\documentclass[12pt]{iopart}
\usepackage[latin9]{inputenc}
\usepackage{amstext}
\usepackage{graphicx}
\usepackage{esint}

\makeatletter
\usepackage{iopams}
\usepackage{setstack}

\usepackage{iopams}\usepackage{setstack}\usepackage{bm}

\makeatother

\begin{document}
\global\long\def\ket#1{\left|#1\right\rangle }

\global\long\def\bra#1{\left\langle #1\right|}

\global\long\def\braket#1#2{\left\langle #1\left|#2\right.\right\rangle }

\global\long\def\ketbra#1#2{\left|#1\right\rangle \left\langle #2\right|}

\global\long\def\braOket#1#2#3{\left\langle #1\left|#2\right|#3\right\rangle }

\global\long\def\mc#1{\mathcal{#1}}

\global\long\def\nrm#1{\left\Vert #1\right\Vert }

\title{The multilevel four-stroke swap engine and its environment}

\author{Raam Uzdin and Ronnie Kosloff }

\address{Fritz Haber Research Center for Molecular Dynamics, Hebrew University
of Jerusalem, Jerusalem 91904, Israel}

\ead{raam@mail.huji.ac.il}

\ead{ronnie@fh.huji.ac.il}

\maketitle
A multilevel four-stroke engine where the thermalization stokes are
generated by unitary collisions with bath particles is analyzed. Our
model is solvable even when the engine operates far from thermal equilibrium
and in the strong system-bath coupling. Necessary operation conditions
for the heat machine to perform as an engine or a refrigerator are
derived. We relate the work and efficiency of the device to local
and non-local statistical properties of the baths (purity, mutual
coincidence etc.). In particular, we relate the Clausius inequality
to the symmetrized relative entropy of the baths (Jefferys divergence).
Other Clausius-like inequalities are derived as well. Finally, in
the ultra-hot regime we optimize the work of the multilevel engine
and obtain simpler forms for the work and efficiency.

\pacs{32.80.Qk, 03.67.-a, 03.65.Yz, 02.30.Yy}

\section{Introduction}

Present day technology is on the verge of enabling different realizations
of quantum heat machines where the engine core (``working substance'')
comprises of a single particle in a discrete level system (in particular
a qubit). In analogy to their classical counterparts quantum heat
machines can be used to cool or to produce work. Furthermore, similarly
to classical thermodynamics, performance analysis of heat machines
without going into the details of a specific realization is a key
theme in quantum thermodynamics (QT). There are several approaches
to QT . The more recent one comes from quantum information resource
theory. In this approach thermal states are considered free while
non-equilibrium states are considered as a resource (\cite{horodecki2013fundamental}
and references therein). An energy conserving unitary is used to couple
the system and the bath. A different viewpoint to QT is dynamical,
and uses the framework of quantum generators of open systems \cite{k281}.
In this approach, models of quantum heat machine are constructed and
analyzed. To a certain extent the present article combine the two
approaches by analyzing the action of a heat engine constructed from
discrete elementary unitary interactions with the reservoirs. A third
approach called typicality aims to find the condition when complex
interactions lead to thermal behavior for a typical set of states
(see \cite{GemmerQTbook} and references therein). 

The goal of the model studied here is to lead to a more comprehensive
understanding of multilevel quantum heat engine operation and to study
the byproducts of their operation. In addition, the open system approach
is typically restricted to a weak coupling to the baths. Even this
weak coupling should be carefully handled in order to avoid conflicts
with the second law \cite{k281}. Our model is both solvable and consistent
with the second law in all coupling strengths.

Various quantum engine models have been suggested - some are continuous
(where the baths are always connected to the engine) and some are
reciprocating (where in different strokes different processes takes
place). The equivalence between a laser and the Carnot engine \cite{scovil59}
has inspired the study of quantum heat engines. For example studying
maximum power operation \cite{k24,k102,abah12,jahnkemahler08}, the
role of coherence and entanglement \cite{scully03,scully11,mukamel12,rahav12,popescu13,popescu13b,dillenschneider2009energetics},
quantum refrigerators and the third law of thermodynamics \cite{k122,k169,k156,k272,segal06,popescu10,segal13}
and the connection to quantum information \cite{kim2011quantum,zhou2010minimal}.

Different quantum working mediums have been considered, two-level
systems, N-level systems or harmonic oscillators \cite{k85,k116,mahler07b,k221}.
In most previous studies the system bath dynamics was modeled by a
reduced description where the dynamics of the system is described
explicitly by a Master equation with a generator cast into the Lindblad
form \cite{lindblad76,kossakowski76}. In the reduced formalism the
bath is only considered implicitly, therefore the effect of the engine
on the baths is almost always ignored. In this paper we take the opposite
point of view and study the heat machine operation using the properties
of the baths. One of the main difficulties in the Lindblad approach
is that the reduced dynamics generators are not uniquely defined.
Various choices of the Lindblad generator lead to the same reduced
dynamics of the working medium but to different entropy fluxes in
the baths. As it turns out the Lindblad operators have to be carefully
chosen in order to be consistent with the second law of thermodynamics
\cite{k281}. The problem becomes more severe when the time dependent
external fields drive the system \cite{k278,LevyKosloff2014local}.

The analysis carried out in this paper overcomes this problem by considering
a very simple yet physically plausible bath model that is based on
collisions \cite{scarani2002thermalizing,ziman2002diluting}. In this
model the bath consists of non-interacting particles which are initially
in the same thermal state. The engine interacts with the bath particles
via a two-particle collision described by a unitary operation. We
choose to model the collision by the unitary swap operation but other
unitary operations such as CNOT and random unitaries have been considered
as well \cite{ziman2005all,gennaro2008entanglement} even though they
were not applied to the study of heat machines. As will be discussed
later on, the swap operation has two interesting physical limits.
The first is the full swap that describes the exchange of particles
and the other is the weak partial swap that mimics quasi-static evolution.

Typically the system that interacts with the bath is taken to be a
single qubit but qudits and a chain of coupled qubits have been considered
as well \cite{GennaroQuDit,burgarth2007mediated}. The collision model
can also represent a general non-Markovian dynamics \cite{rybar2012simulation,bodor2013structural}.
If, however, the particle that interacted with the system is extremely
unlikely to interact again with the system or the bath particles,
then the dynamics is Markovian. See \cite{ziman2005description,ziman2003saturation}
and reference therein for more studies on collision-generated Markovian
dynamics. In this article we consider this type of Markovian collisions
in order to study the most basic features of an engine driven by collision
baths.

The importance of studying multilevel heat machine is two-fold. First,
experimentally speaking, in some systems it is not possible or not
practical to interact with only two levels. Secondly, two-level systems
may contain some non-generic features and it important to understand
the more general framework of quantum thermodynamics. For example,
in a two-level system temperature is always well defined if the density
matrix is diagonal in the energy basis. Clearly this is not true if
there are more than two levels.

While in a two-level engine it is fairly simple to write down the
map the parameter space where the system operates as an engine and
as refrigerator, this situation is considerable more complicated when
there multiple levels. The different operation regimes are determined
by the temperature and the gap energy associated with the cold and
hot strokes. It is not straightforward what energy scale replaces
the two-level energy gap in the multilevel case (the variance? the
maximal gap?). Furthermore it is not clear at this point if necessary
and sufficient conditions for engine (or refrigerator) operation can
be formulated without explicitly solving the full dynamics of the
system.

The present analysis can be related to ``finite time thermodynamics''
\cite{salamon01,curzon75}although there is no explicit reference
to time. Implicitly the cycle can be related to the collision timscale
and in addition there is no assumption of thermal equilibration of
the device with the baths upon contact.

The article is organized as follows: Section 2 contains the description
of our engine and bath model. In Sec. 3 we analyze the steady state
operation of the engine and discuss the difference between quantum
swap and classical random swap. Next, Sec. 4 explores various aspects
of the engine evolution: Clausius and, generalized Clausius inequality
and the baths purity reduction. In Sec. 5 we set upper bounds on the
work and efficiency in term of statistical quantities like purity,
mutual coincidence and Wooters distance. Finally, Sec. 6 presents
two possible regimes of operation: the ultra-hot bath regime and the
quasi-static regime.

\section{Baths, engine, and their interaction}

\subsection{The bath}

Our bath model is based on the ``quantum homogenizer'' introduced
in Ref. \cite{scarani2002thermalizing,ziman2002diluting,diosi2006exact}.
This bath contains a very large amount of single species non-interacting
multilevel particles. At the beginning all the particles are in the
same thermal state. However, each time a bath particle interact with
the engine the population of the particle and of the engine changes
according to the simple partial swap rule that will be described shortly. 

Every time the system (engine) interacts with the bath it collides
with a \textit{new }thermal particle in the bath. If the system is
coupled only to one bath then repeated collision would lead to thermalization
of the engine particle at as exponential rate. In this work, however,
we will assume that there is only one collision or none at all at
each thermal stroke of the engine. Therefore, the engine will typically
not be in a Gibbs state. The extension to multiple collisions is straightforward. 

The bath is assumed to be large enough so that the probability to
re-collide with a bath particle that already interacted with the system
is negligible. Consequently, the resulting dynamics is Markovian.
We assume that the particles do no interact with each other in the
bath but only with the system particle at the bath-system interface.
That is, in our model there is no thermalization inside the bath.
In principle, for large baths, this has no impact on the engine's
operation. If the bath is large enough it does not matter if the scattered
particle get thermalized again or not as the engine will (almost)
always interact with a new thermal particle. Nevertheless, it is interesting
to explore the operation of small baths with and without intra-bath
thermalization. 

There is a different class of quantum engines where the baths are
always connected to the engine but different levels are connected
to different baths \cite{k122}. This implies that if only one bath
is connected the \textit{relative} probabilities of the specific coupled
levels will be thermal but the engine as a whole will not be in a
Gibbs state. Alternatively stated, these bath models have a continuum
of steady states. In our model the Gibbs state is the only single
bath steady state.

\subsection{The engine's cycle}

Our engine comprises of a multilevel system that is driven to a four
stroke Otto cycle \cite{k152}. Figure 1a illustrate the engines operation
for a two-level system. In stroke $A$ the gap spacing is increased
by applying an external field. The engine is decoupled from the baths
at this stage. The state evolution is such that the system's density
is diagonal in the energy basis at the beginning and at the end of
the stroke and the populations do not change. Adiabatic change of
the Hamiltonian will achieve such a transformation but other faster
options exist (see Sec. \ref{sub: The-adiabatic-evolution}). Nonetheless,
this part of the cycle is termed ``adiabatic''. 

In stroke $B$ the gap is kept fixed in time, but the engine is allowed
to interact with the hot bath. During this time interval there may
be one collision, multiple collisions or none at all. The average
collision rate is a parameter that characterizes the bath and it's
coupling to the engine. It encapsulates within the particle density
in the bath, the engine cross-section for collision etc. The collision
entangles the engine and the particles in the bath. However since
the interacting bath particles will not interact with the the engine
again (the top spheres with temperature denoted by primes), we consider
the reduced dynamics of the engine by taking a partial trace over
the scattered particle of the bath. This stage changes the entropy
of the engine. In stroke $C$ the external field is changed again
so that the gap returns to its initial value. Finally, in stroke $D$,
the system is coupled to the cold bath.

In the multilevel case the evolution can be considerable more intricate.
Firstly, we do not assume that all the levels are increase or decreased
by the same ratio (Fig. 1c). Secondly, we allow the levels to cross
(Fig. 1d). \\
\begin{figure}
\includegraphics[width=16cm]{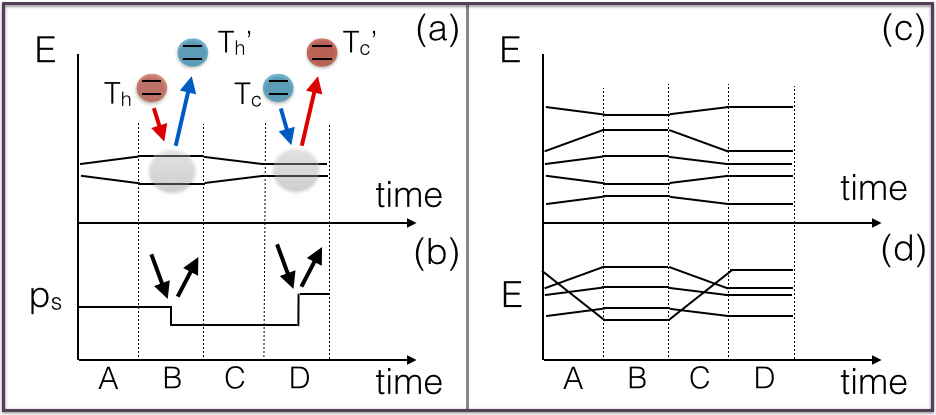}

\caption{(1a) The energy of the ground state and excited state of the two level
system that comprises four-stroke, two-level, partial swap collision
driven engine. (1b) Dynamic of the ground state population. In strokes
$A$ and $B$ the population is fixed and the energy gap changes and
in stroke $C$ and $D$ the gap is fixed and the population changes
via collisions with the baths particles. \textit{In general the collision
stage is not an isotherm}. The work is done during stroke $A$ and
$C$ and heat exchange takes place during strokes $B$ and $D$. (1c)
In a multilevel the level dynamics can be more intricate. The levels
do not have to be compressed by the same factor. Furthermore they
can even cross each other (1d) as long as it does not conflict with
energy population invariance in stages $B$ and $C$. }
\end{figure}

\subsection{The collision model}

We assume that at the initial state the system and the bath particles
are in a product of thermal Gibbs state: 
\begin{eqnarray}
\rho & = & \rho_{b}\otimes\rho_{b}\otimes\rho_{b}....,\\
\rho_{b,ii} & = & e^{-\frac{E_{b,i}}{T_{b}}}/\sum_{j=1}^{N}e^{-\frac{E_{b,i}}{T_{b}}}=e^{-\frac{E_{b,i}-F_{b}}{T_{b}}},\\
\rho_{b,i\neq j} & = & 0,
\end{eqnarray}
where '$b$' stands for '$c$' (cold bath) or '$h$' (hot bath). The
free energy $F_{b}$ takes care of normalization. $ $Let U be some
two-particle unitary operation that conserves the total energy of
the two particles. The reduced density matrices after the interaction
are:
\begin{eqnarray}
\rho_{b}' & = & tr_{s}(U\rho_{b}\otimes\rho_{s}U^{\dagger}),\\
\rho_{s}' & = & tr_{b}(U\rho_{b}\otimes\rho_{s}U^{\dagger}).
\end{eqnarray}
\[
\]
When the collision Hamiltonian that generates U is invariant under
the transformation $b\leftrightarrow s$ one can show that:
\begin{equation}
\rho_{b}'-\rho_{b}=-(\rho_{s}'-\rho_{s}).\label{eq: dp -dp}
\end{equation}
This condition, together with the total energy conservation, implies
that the energy levels of the bath and the system have to be equal.

\subsection{Density matrix swap and energy population swap}

In this work we focus on collision that induce a single parameter
convex transformation for reduce density matrices or for the energy
population. The density swap rule is:
\begin{eqnarray}
\rho_{s}' & = & (1-x)\rho_{s}+x\rho_{b},\label{eq: rho_s swap}\\
\rho_{b}' & = & (1-x)\rho_{b}+x\rho_{s},\label{eq: rho_b swap}\\
0 & \le x\le & 1,\label{eq: x domain}
\end{eqnarray}
where $x$ is the swap parameter. For $x=0$ nothing happens and for
$x=1$ the density matrix of the bath and the particle completely
swap. It is easy to verify that this transformation satisfies (\ref{eq: dp -dp})
and that the only steady state of this transformation is $\rho_{b}=\rho_{s}$. 

Re-colliding the system particles with new bath particles in a thermal
state will lead to exponential decay of $\rho_{s}$ towards $\rho_{b}$.
Yet, in this work, we do not assume that equilibrium is reached. In
fact we will mostly consider a single collision at most in a given
stroke ($B$ or $D$). 

Although the density swap rule has a very simple structure our framework
is valid for a more general type of swap operation. Let $\rho_{b}$
and $\rho_{s}$ be diagonal matrices in the energy basis, where $\mathbf{p}_{b}$
and $\mathbf{p}_{s}$ denote the energy level populations (the diagonal
terms in the density matrix) of the bath and system particle. We shall
use bold face characters to denote quantities that have a level index
such as probability or energies of a bath $\mathbf{E}_{b}$. The energy
population swap rule is:

\begin{eqnarray}
\mathbf{p}_{s}' & = & (1-x)\mathbf{p}_{s}+x\mathbf{p}_{b},\label{eq: p_s swap}\\
\mathbf{p}_{b}' & = & (1-x)\mathbf{p}_{b}+x\mathbf{p}_{s},\label{eq: p_b swap}\\
0 & \le x\le & 1,\label{eq: x p  domain}
\end{eqnarray}
where as before the prime relates to the traced-out outcome after
the collision. We do not claim that the swap operation in its present
form is general enough to describe many typical thermal baths used
in experiments. Yet, such bath could in principle be built. One advantage
of this bath model is that it leads to a solvable dynamics that can
be used as a reference for other more complex model. Another advantage,
as will be shown later on, is that it helps to formulate new questions
and new points of view concerning the operation of heat machines.
Nevertheless, in a two-level system an energy population swap follows
immediately from (\ref{eq: dp -dp}) since only one parameter is needed
to describe the population transfer in a two-level system. For example,
the Einstein rate equation for an interaction of two levels with thermal
radiation describes such an energy population swap interaction.

In appendix I, we show examples of two (partial) swap Hamiltonians.
One generates a density matrix swap and the other an energy population
swap.

\subsection{The adiabatic evolution step\label{sub: The-adiabatic-evolution}}

Stroke $A$ and $C$ are termed ``adiabatic'' as we require that
the evolution will be diagonal in the energy basis at the beginning
and the end of the stroke. The coherences will remain zero and the
energy populations will remain as they are. Such a process does not
have to be slow. 

For example, consider the Hamiltonian that commutes with itself at
all times such as:
\begin{equation}
H_{adiab}=\sum_{i=1}^{N}E_{i}(t)\ketbra ii.\label{eq: H adiab}
\end{equation}
A density matrix that is diagonal in the energy basis $\ket i$ will
be invariant under this type of Hamiltonian regardless of how fast
the energy levels are changing. In a two-level spin system this Hamiltonian
will be $H_{adiab}=B_{z}(t)\sigma_{z}$, where $\sigma_{z}$ is the
$z$ Pauli matrix.

The population and coherences evolution can be obtained by more general
time dependent Hamiltonians that do not commute at different times
$[H(t_{1}),H(t_{2}\neq t_{1})]\neq0$. Nevertheless the possibility
of an adiabatic transformation is guaranteed by the coherent control
theorem \cite{huang1983controllability,ramakrishna1995controllability}.
Alternatively it is possible to use a method known as ``quantum driving''
or ``shortcut to adiabaticity'' \cite{demirplak2003adiabatic,demirplak2008consistency,berry2009transitionless,chen2011optimal}
to generate an evolution that preserves the energy-basis diagonal
form of the density matrix at the end of the process. 

In principle, we allow the energy levels to cross%
\footnote{this will not automatically generate strong non-adiabatic effects
(e.g. if (\ref{eq: H adiab}) drives the system).%
}. However in such a case it important to remember that the energy
index is a level index and not some order index that indicates how
the levels are ordered. For example, in Fig 1b-buttom $E_{c,1}<E_{c,2}<E_{c,3}<E_{c,4}$
but at the hot bath $E_{h,4}<E_{h,1}<E_{h,2}<E_{h,3}$.

\section{\label{sec: evo fluct}The average populations in steady state operation}

The dynamics of the system (engine) reduced density $\rho_{s}$ and
its statistical features are explored. We assume that during stroke
$B$ a single particle interact with the engine with probability $R$.
The description can easily be generalized to include the possibility
of more than one collision during stroke $B$ or $D$. The probability
$R$ appears naturally in a collisions model as there is no certainty
that a particle from the a bath particle will be available to interact
with the engine. For simplicity we assume that the bath parameters
are identical for both baths: they have same $R$ and the same swap
parameter $x$. A generalization to different x and R for different
baths is straightforward using the same methods.

Even though for some applications transient behavior may be of interest,
here we focus on the steady state operation. The average engine population
at stage $C$, $\rho_{s}^{C}$, is related to that of stage $A$,
$\rho_{s}^{A}$, via:

\begin{equation}
\left\langle \rho_{s}^{C}\right\rangle =(1-R)\left\langle \rho_{s}^{A}\right\rangle +R[(1-x)\left\langle \rho_{s}^{A}\right\rangle +x\rho_{h}].\label{eq: rho basic C to A}
\end{equation}
The first term describes a no-collision event and the other describes
a collision with a swap parameter x. Equation (\ref{eq: rho basic C to A})
simplifies to:
\begin{equation}
\left\langle \rho_{e}^{C}\right\rangle =(1-xR)\left\langle \rho_{e}^{A}\right\rangle +xR\rho_{h}.\label{eq: rho e C}
\end{equation}
Here $x$ and $R$ are inseparable since both of them have the same
effect on the average population. In the same manner we can write
equation for $\rho_{s}^{A}$:
\begin{equation}
\left\langle \rho_{s}^{A}\right\rangle =(1-xR)\left\langle \rho_{s}^{C}\right\rangle +xR\rho_{c}.\label{eq: rho e A}
\end{equation}
By combing (\ref{eq: rho e C}) and (\ref{eq: rho e A}) we get: 
\begin{eqnarray}
\left\langle \rho_{s}^{C}\right\rangle  & = & \frac{\rho_{h}+\rho_{c}-xR\rho_{c}}{2-xR},\label{eq: rho C steady}\\
\left\langle \rho_{s}^{A}\right\rangle  & = & \frac{\rho_{h}+\rho_{c}-xR\rho_{h}}{2-xR}.\label{eq: rho A steady}
\end{eqnarray}
Note that a combination of Gibbs states is not a thermal state if
there are more than two levels. Hence, as expected in a finite time
thermodynamic framework, the multilevel swap engine is never in a
Gibbs state for $xR<1$. The expectation value of the population change
is:
\begin{equation}
d\rho_{s}^{A\to C}=\left\langle \rho_{s}^{C}\right\rangle -\left\langle \rho_{s}^{A}\right\rangle =\frac{xR}{2-xR}(\rho_{h}-\rho_{c}).
\end{equation}
Inspection of (\ref{eq: rho C steady}) and (\ref{eq: rho A steady})
shows that even if the initial density matrix of the engine has some
coherences they have no impact on the energy diagonal steady state.
Thus for all energy observables computed locally (i.e., with the reduced
density matrices of the baths or the engine) it is sufficient to consider
the population vector $\mathbf{p}_{i}=diag(\rho_{i})$ where $i='s','c','h'$.
Hence, in this notation:
\begin{equation}
d\mathbf{p}_{e}^{A\to C}=\frac{xR}{2-xR}(\mathbf{p}_{h}-\mathbf{p}_{c}).\label{eq: dp avg}
\end{equation}
In general, the population change under a swap operation in steady
state is always proportional to $\mathbf{p}_{h}-\mathbf{p}_{c}$.
This result will have a large impact later on. We point out that if
the baths are only coupled to some of the system levels as in \cite{scovil59},
then the result is different. 

When $xR=1$ in a single collision, or $xR<1$ with a vast number
of collisions a complete thermalization of the engine's population
takes place. Therefore, many of the results in the limit $xR=1$,
apply to a much more general setup than the swap model described above.
They apply wherever the engine is connected long enough to effectively
reach a Gibbs state (and not some other state). 

For work and efficiency investigations, only the populations in the
energy basis are important. Hence, it is not necessary that the whole
density matrix is swapped as in (\ref{eq: rho_s swap})-(\ref{eq: x domain}),
rather it is sufficient that an energy population swap takes place
(see Eqs. (\ref{eq: p_s swap})-(\ref{eq: x p  domain})).

\subsection{Quantum vs. classical swap and entropy generation}

In Eq. (\ref{eq: dp avg}) $x$ and $R$ are lumped together in a
product form $xR$. Yet, they are not physically equivalent. $R$
controls how deterministic the system is, and $x$ determines the
interaction strength and the ``quantumness'' (coherences and entanglement
before the partial trace). Note that the quantum behavior is not monotonically
increasing with $x$, since the system is classical when $x=1$. 

In the quantum case, where $R=1$ $x<1$, $x$ can generate coherences
and entanglement in the joint density matrix of two interacting particles.
The entropy increase in the sum of entropies of the reduced density
matrices is the result of ignoring entanglement and classical correlations. 

In contrast, in the classical case where $x=1$, and $R<1$ , the
particles are either fully swapped or left as they are. Therefore
entanglement cannot be produced from the initial product state. Here,
the sum of the entropies of the individual particles also increases
since the information if a collision took place or not is discarded. 

In both cases the total entropy increase of the reduced entropies
is encapsulated in the mutual information of the colliding particles.
In the quantum case, the mutual information contains contribution
from entanglement. The separation to quantum and classical correlations
can be studied using the quantum discord tool \cite{ollivier2001quantumDiscord}.
This however is outside the scope of this work. Furthermore in our
model the observables we studied depend only on $xR$ so they can
equally be obtained from a classical or a quantum realization. However,
we do no expect this to hold for observables that are not functions
of the steady state population.

\section{Thermodynamic properties}

\subsection{First law\label{sec: 1st 2nd laws}}

In the present model the expectation value of the energy is: 
\begin{eqnarray}
\frac{dU}{dt} & = & \frac{d}{dt}(\mathbf{p}_{s}\cdot\mathbf{E})=\\
 & = & \underset{\frac{d}{dt}Heat}{\underbrace{\mathbf{E}\cdot\frac{d}{dt}\mathbf{p}_{s}}}+\underset{-\frac{d}{dt}Work}{\underbrace{\mathbf{p}_{s}\cdot\frac{d}{dt}\mathbf{E}}}.
\end{eqnarray}
On average in a complete cycle $\left\langle U\right\rangle _{initial}=\left\langle U\right\rangle _{final}$
we have: 
\begin{eqnarray}
0 & = & \left\langle \mathbf{p}_{s}^{A}\right\rangle (\mathbf{E}_{h}-\mathbf{E}_{c})+\mathbf{E}_{h}\cdot(\left\langle \mathbf{p}_{s}^{C}\right\rangle -\left\langle \mathbf{p}_{s}^{A}\right\rangle )\nonumber \\
 &  & +\left\langle \mathbf{p}_{s}^{C}\right\rangle \cdot(\mathbf{E}_{c}-\mathbf{E}_{h})+\mathbf{E}_{c}\cdot(\left\langle \mathbf{p}_{s}^{A}\right\rangle -\left\langle \mathbf{p}_{s}^{C}\right\rangle ),
\end{eqnarray}
where the first term correspond to stroke $A$, the second to $B$,
and so forth. Regrouping we identify:
\begin{eqnarray}
\left\langle Q_{h}\right\rangle  & = & \mathbf{E}_{h}\cdot\left\langle d\mathbf{p}_{s}^{A\to C}\right\rangle ,\label{eq: Qh def}\\
\left\langle Q_{c}\right\rangle  & =- & \mathbf{E}_{c}\cdot\left\langle d\mathbf{p}_{s}^{A\to C}\right\rangle ,\\
\left\langle W\right\rangle  & = & \left\langle d\mathbf{p}_{s}^{A\to C}\right\rangle \cdot(\mathbf{E}_{h}-\mathbf{E}_{c}),\label{eq: work def}
\end{eqnarray}
where we used the ``positive heat in, positive work out'' sign convention.
Note that these quantities are invariant to any constant shift of
the level in one or two of the bath: $E_{h,i}\to E_{h,i}+F_{h},E_{c,i}\to E_{c,i}+F_{c}$
where $F_{c}$ and $F_{h}$ are some constants. The probabilities
themselves are invariant to such transformation by virtue of the exact
form of the Gibbs state. The energy and heat are not invariant to
such a shift but the extra term cancels out when summed over the probability
difference. This is true for the steady state heat flow to each bath.
Equations (\ref{eq: Qh def})-(\ref{eq: work def}) lead to the averaged
form of the first law:
\begin{equation}
\left\langle Q_{h}\right\rangle +\left\langle Q_{c}\right\rangle =\left\langle W\right\rangle .
\end{equation}
Note that the energy at the end of the cycle is equal to the energy
at the beginning of the cycle only on average. In a specific cycle
it is not zero. Since the engine's Hamiltonian is time dependent its
energy need not be conserved at each instant. Using (\ref{eq: dp avg})
and (\ref{eq: work def}) we obtain:
\begin{equation}
\left\langle W\right\rangle =\frac{xR}{2-xR}(\mathbf{p}_{h}-\mathbf{p}_{c})\cdot(\mathbf{E}_{h}-\mathbf{E}_{c}).\label{eq: <W>}
\end{equation}
\[
\]

\begin{figure}
\includegraphics[width=10cm]{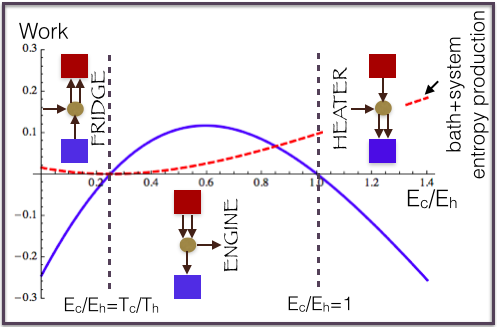}

\caption{\label{fig: work entropy}A typical work per cycle curve (blue-solid)
of a two-level partial swap heat machine as a function of the cold
bath energy gap, $\Delta E_{c}$, when $\Delta E_{h}$ is held fixed.
Below the Carnot point (left vertical line, $\Delta E_{c}/\Delta E_{h}=T_{c}/T_{h}$)
the device acts as a refrigerator and above it until $\Delta E_{c}=\Delta E_{h}$
it performs as an engine. $ $For $\Delta E_{c}>\Delta E_{h}$ the
device performs as a heater as it takes work to make the cold bath
hotter. The dashed-red curve shows the total entropy of the baths
and system. }
\end{figure}
As mentioned before, the limit $xR=1$, (\ref{eq: <W>}) holds for
any interaction that leads to a complete thermalization (or very close
to it) and not just for a swap interaction. The modes of operation
of a two-level swap machine are shown in Fig. \ref{fig: work entropy}.
In general it is not straightforward to write analytically the condition
for engine or refrigerator operation using the energy levels and the
temperature due to the exponential dependence of the energy difference
on the same parameters. Yet, in some temperature regimes this is considerably
simpler as discussed in section \ref{sub: Ultra hot regime}.

\subsection{Second law and the Clausius number\label{sub:Second-law-and} }

In cyclic processes in classical thermodynamics the second law can
be expressed in terms of the Clausius inequality $\oint\frac{\delta Q}{T}\ge0$.
In our case we calculate:
\begin{equation}
\mc R=\frac{\left\langle \Delta Q_{h}\right\rangle }{T_{h}}+\frac{\left\langle \Delta Q_{c}\right\rangle }{T_{c}},\label{eq: C no. def}
\end{equation}
where, as before, the bracket denotes the average value in steady
state. Strictly speaking, this is not the classical Clausius inequality.
The swap collision process is not an isotherm at all. For a multilevel
system a temperature cannot be assigned to the bath or engine particles
since they are not in a Gibbs state after (or during) the collision.
In this section we find that $\mc R\ge0$ holds in steady state for
any multilevel swap engine and that it has an information theoretic
interpretation. Furthermore, it is shown that $\mc R\ge0$ belongs
to a family of more general inequalities. It is easy to verify that
$\mc R\ge0$ entails within the second law. For $T_{c}=T_{h}$ we
get $W\le0$. That is, \textit{in steady state}, no work can be extracted
from a single bath. Alternatively, when calculation the efficiency,
$\eta=1-\left|\frac{\left\langle Q_{c}\right\rangle }{\left\langle Q_{h}\right\rangle }\right|$
, $\mc R\ge0$ ensures that the efficiency is smaller than the Carnot
efficiency (see Sec. \ref{sub: efficiency bounds}). 

Using the expressions for heat we want to show that for swap heat
machines:
\begin{equation}
\mc R=\sum_{i=1}^{N}\left\langle dp_{i}^{A\to C}\right\rangle (\frac{E_{h,i}}{T_{h}}-\frac{E_{c,i}}{T_{c}})\ge0.\label{eq: C no. dp > 0}
\end{equation}
In a specific cycle this does not have to be true. Our aim is to show
that it holds on average when the system is in steady state. To obtain
the average we use (\ref{eq: dp avg}) and get:

\begin{eqnarray}
\mc R & = & \frac{xR}{2-xR}\sum_{i}(p_{c}-p_{h})(\frac{E_{h,i}}{T_{h}}-\frac{E_{c,i}}{T_{c}})\\
 & = & \frac{xR}{2-xR}\sum_{i}p_{c,i}\ln\frac{p_{c,i}}{p_{h,i}}+p_{h,i}\ln\frac{p_{h,i}}{p_{c,i}},
\end{eqnarray}
which leads to the following result for swap engines:
\begin{eqnarray}
\mc R & = & \frac{xR}{2-xR}J(\mathbf{p}_{c},\mathbf{p}_{h}),\\
J & = & D_{KL}(\mathbf{p}_{c}|\mathbf{p}_{h})+D_{KL}(\mathbf{p}_{h}|\mathbf{p}_{c}),
\end{eqnarray}
where $J$ is the Jefferys divergence and $D_{KL}(\mathbf{p}|\mathbf{q})=\sum_{i}p_{i}\ln\frac{p_{i}}{q_{i}}$
is the Kullback-Leibler divergence or the relative entropy. A similar
result is known for isothermal processes \cite{DeffnerLutz2010generalizedClausius}.
Since $D_{KL}(\mathbf{p}|\mathbf{q})\geq0$ for any two probability
vectors (also known as Klein's inequality), $\mc R\ge0$ follows naturally
as information theory inequality.

\subsection{Generalized Clausius inequality}

In this section we show that the Clausius number inequality is a special
case of a family of inequalities that hold for swap heat machines:
\begin{equation}
\mc R_{2m-1}=\sum_{i=1}^{N}\left\langle dp_{i}^{A\to C}\right\rangle (\frac{E_{h,i}}{T_{h}}-\frac{E_{c,i}}{T_{c}})^{2m-1}\ge0.\label{eq: R 2m-1}
\end{equation}
Clearly, if this equality holds, it holds for any odd analytic and
monotonically increasing function of $\frac{E_{h,i}}{T_{h}}-\frac{E_{c,i}}{T_{c}}$
as well. The proof is straightforward. While $\mc R\equiv\mc R_{1}$
can be understood in thermodynamics terms of heat, temperature and
entropy, for $m>1$ higher energy powers are involved and there is
no straightforward thermodynamic interpretation. $ $

Denoting $\varepsilon_{b,i}=E_{b,i}/T_{b}$ and $D_{i}=\varepsilon_{h,i}-\varepsilon_{c,i}$
we get that inequality (\ref{eq: R 2m-1}) is equivalent to:
\begin{equation}
\mc R_{2m-1}=\frac{\frac{xR}{2-xR}}{\sum_{k}e^{-\varepsilon_{c,k}}\sum_{k'}e^{-\varepsilon_{h,k'}}}\sum_{ij}e^{-\varepsilon_{c,i}-\varepsilon_{c,j}}(e^{-D_{j}}-e^{-D_{i}})D_{i}^{2m-1}.
\end{equation}
Next we write $\mc R_{2m-1}=\frac{1}{2}\mc R_{2m-1}+\frac{1}{2}\mc R_{2m-1}$,
exchange the indexes names in the second term and obtain: 
\begin{eqnarray}
\mc R_{2m-1} & = & N_{b}\frac{1}{2}\frac{\frac{xR}{2-xR}}{\sum_{k}e^{-\varepsilon_{c,k}}\sum_{k'}e^{-\varepsilon_{h,k'}}}\times\nonumber \\
 &  & \sum_{ij}e^{-\varepsilon_{c,i}-\varepsilon_{c,j}}(e^{-D_{j}}-e^{-D_{i}})(D_{i}^{2m-1}-D_{j}^{2m-1}).\label{eq: C 2m-1}
\end{eqnarray}
The term $D_{i}^{2m-1}-D_{j}^{2m-1}$ has the same sign as $D_{i}-D_{j}$
and the same sign as $(e^{-D_{j}}-e^{-D_{i}})$, hence the product
of the two differences that appears in (\ref{eq: C 2m-1}) is always
positive. The rest of the multipliers are positive and symmetric under
$i\leftrightarrow j$ and therefore $\mc R_{2m-1}\ge0$.

\subsection{Clausius dominated level}

Immediate consequence of this generalized Clausius number inequality
follows from considering $m\to\infty$. In this case, the term with
largest $\left|\frac{E_{h,i}}{T_{h}}-\frac{E_{c,i}}{T_{c}}\right|$
becomes enormously larger than all the other terms. Therefore, this
single term in $\mc R_{\infty}$ must be positive. If this term is
positive, it is positive also in $\mc R_{1}$ and therefore the Clausius
dominated level $i_{max}$ defined by:
\begin{equation}
\left|\frac{E_{h,i_{max}}}{T_{h}}-\frac{E_{c,i_{max}}}{T_{c}}\right|\ge\left|\frac{E_{h,i}}{T_{h}}-\frac{E_{c,i}}{T_{c}}\right|,
\end{equation}
satisfy:

\begin{equation}
\text{sign}\left\langle dp_{i_{max}}^{A\to C}\right\rangle =\text{sign}(\frac{E_{h,i_{max}}}{T_{h}}-\frac{E_{c,i_{max}}}{T_{c}}),\label{eq: sgn C dominated}
\end{equation}
where we assume that $\left|\frac{E_{h,i}}{T_{h}}-\frac{E_{c,i}}{T_{c}}\right|$
has a single maximum and that $\left\langle dp_{i_{max}}^{A\to C}\right\rangle \neq0$.
Equation (\ref{eq: sgn C dominated}) allows us to immediately determine
the direction of heat flow into the baths for this level. Of course,
this can be done explicitly by evaluating numerically the probabilities.
This equality is not trivial since the Clausius dominated level is
not necessarily the largest element in the Clausius number sum (\ref{eq: C no. dp > 0}).

\subsection{Local and non-local quantities}

We call a quantity 'local scalar' if the scalar quantity can be written
in terms of the single-bath parameters $\mathbf{E}_{b}$ and $T_{b}$
(i.e. $f(\mathbf{E}_{b},T_{b})$). The mean energy $\mathbf{p}_{b}\cdot\mathbf{E}_{b}$
of the bath particle, its purity $\mathbf{p}_{b}\cdot\mathbf{p}_{b}$,
its entropy, its energy variance and its free energy, are all examples
local scalars. A local function is a function of local scalars. 

Many other quantities of prime importance cannot be written in this
form: the mutual coincidence $\mathbf{p}_{c}\cdot\mathbf{p}_{h}$,
the fidelity of the two baths $\sum_{i}\sqrt{p_{c,i}}\sqrt{p_{h,i}}$,
the Jefferys and Kullback-Leibler divergence, the Wootters distance.
Hence these scalars are 'non-local scalars'. \textit{More importantly
heat, work and efficiency are non-local scalars}. 

\textit{In our article thermodynamics amounts to finding relations
between non-local scalar (e.g. work and efficiency) and local scalars
(temperature entropy etc.). }

\subsection{Purity reduction}

Let us define the purity of the bath after the collision by the purity
of the reduced density formed by tracing out the engine and the other
bath. Although the purity does not have all the appealing properties
of the von Neumann entropy, it provides a simple and convenient measure
of impurity. Furthermore, in contrast to entropy, it is not sensitive
to whether we keep track of the particles position or not (i.e. there
is no mixing entropy issue). In addition the purity will naturally
emerge in the derivation of bound on maximal work and efficiency.

The purity change in one cycle in one bath is:
\begin{equation}
\Delta\mc P_{b}=\left|\mathbf{p}_{b}+d\mathbf{p}_{b}\right|^{2}-\left|\mathbf{p}_{b}\right|^{2}=d\mathbf{p}_{b}^{2}+2\mathbf{p}_{b}\cdot d\mathbf{p}_{b},
\end{equation}
where the absolute value of a vectors refers to the standard $L_{2}$
norm $\left|\mathbf{y}\right|=\sqrt{\sum_{i=1}^{N}\left|y_{i}\right|^{2}}\equiv\sqrt{\mathbf{y}^{2}}$.

Using $d\mathbf{p}_{c}=-d\mathbf{p}_{h}$ for steady state operation
we finally obtain:
\begin{eqnarray}
\Delta\mc P_{h}+\Delta\mc P_{c} & = & 2(\frac{1-xR}{xR})d\mathbf{p}_{h}^{2}\\
 & =- & \frac{xR(1-xR)}{(2-xR)^{2}}\left|\mathbf{p}_{h}-\mathbf{p}_{c}\right|^{2}.
\end{eqnarray}
In a refrigerator the cold bath the purity increases while the hot
bath becomes more mixed. Yet, the change in the hot bath is larger
so the purity of the whole system decreases. For $xR<1$ the whole
system (engine+baths) becomes increasingly mixed in all modes of operations.

In the spirit of section we want to express in the purity reduction
using local functions. This can be achieved by applying the inverse
triangular inequality:
\begin{eqnarray}
\left|\Delta\mc P_{h}+\Delta\mc P_{c}\right| & \ge & \frac{xR(1-xR)}{(2-xR)^{2}}\left|\left|\mathbf{p}_{h}\right|-\left|\mathbf{p}_{c}\right|\right|^{2}\\
 & = & \frac{xR(1-xR)}{(2-xR)^{2}}\left|\sqrt{\mc P_{h}}-\sqrt{\mc P_{c}}\right|^{2}.
\end{eqnarray}
Interestingly, in contrast to the Clausius inequality, we did not
assume a thermal distribution. In fact, in steady state this device
will decrease the total purity for any bath population that is diagonal
in the energy basis. While the Clausius is not well defined (there
is no notion of temperature for non thermal baths) the purity decrease
still holds. 

Finally, we note that using the Jensen inequality for the $\ln$ function
it is straightforward to show that the purity is related to the entropy
through:
\begin{equation}
\mc P\ge e^{-S},\label{eq: puriy S ineq}
\end{equation}
where $S=-\sum_{i}p_{i}\log p_{i}$ is the standard entropy function.
This inequality can be interpreted in the following way: the Chebyshev
sum inequality yields: $\mc P\ge1/N$ where the equality holds for
uniform distributions. $e^{S}$ is the Shanon's effective number of
degrees of freedom. The right hand side of (\ref{eq: puriy S ineq})
is also equal $1/N$ for uniform distribution. Hence $ $(\ref{eq: puriy S ineq})
expresses the relation between two measures that count the degrees
of freedom in the system.

\section{Mutual coincidence necessary conditions for engines and refrigerators}

The engine regime is defined by the condition $\left\langle W\right\rangle =\frac{xR}{2-xR}(\mathbf{p}_{h}\cdot\mathbf{E}_{h}+\mathbf{p}_{c}\cdot\mathbf{E}_{c}-\mathbf{p}_{c}\cdot\mathbf{E}_{h}-\mathbf{p}_{h}\cdot\mathbf{E}_{c})\ge0$.
Using the mean bath energy $\left\langle E_{b}\right\rangle =\mathbf{p}_{b}\cdot\mathbf{E}_{b}$
and the free energy defined through $p_{b}=e^{-\frac{E_{b,i}-F_{b}}{T_{b}}}$
we get:

\begin{equation}
\left\langle W\right\rangle =\left\langle E_{h}\right\rangle +\left\langle E_{c}\right\rangle +\sum_{i}p_{c,i}(T_{h}\ln p_{h,i}-F_{h})+p_{h,i}(T_{c}\ln p_{c,i}-F_{c}).
\end{equation}
After rearranging and using the Jensen inequality to get $\ln\mathbf{p}_{c}\cdot\mathbf{p}_{h}\ge\sum_{i}p_{c,i}\ln p_{h,i},\ln\vec{p}_{c}\cdot\vec{p}_{h}\ge\sum_{i}p_{h,i}\ln p_{c,i}$
we get that the work is smaller than: 
\begin{equation}
T_{c}S_{c}+T_{h}S_{h}+(T_{h}+T_{c})\ln\mathbf{p}_{h}\cdot\mathbf{p}_{c}\ge\left\langle W\right\rangle ,\label{eq: W engine cond}
\end{equation}
where we used: $T_{b}S_{b}=\left\langle E_{b}\right\rangle -F_{b}$.
If the left hand side is larger than zero so will $\left\langle W\right\rangle $.
Thus, demanding that the left hand side of (\ref{eq: W engine cond})
is positive we get a necessary (but not sufficient) condition for
the heat machine to perform as an engine: 

\begin{equation}
\mc P_{ch}\ge Exp[-\frac{T_{c}S_{c}+T_{h}S_{h}}{T_{c}+T_{h}}],\label{eq: Pch engine cond}
\end{equation}
where $\mc P_{ch}$ is the mutual coincidence $\mathbf{p}_{c}\cdot\mathbf{p}_{h}$.
While the purity describes the probability that the same result will
appear when measuring the energy of two particles in the same bath,
$\mc P_{ch}$ describes the probability that particles from different
baths will be in the same level. $\mc P_{ch}$ is often used in cryptography
and code ciphering. Note, that it does not imply a correlation but
simply the likelihood of identical events in both baths. 

Note that (\ref{eq: Pch engine cond}) actually imposes a restriction
also on the individual purities since:
\begin{equation}
\frac{\mc P_{c}+\mc P_{h}}{2}\ge\sqrt{\mc P_{c}\mc P_{h}}\ge\mathbf{p}_{c}\cdot\mathbf{p}_{h}\ge e^{-\frac{T_{c}S_{c}+T_{h}S_{h}}{T_{c}+T_{h}}}.
\end{equation}
For the refrigerator the work is not a good indicator since it is
possible to apply work to the system without cooling the cold bath.
Cooling occurs when the cold bath gives away heat to the system: 

\begin{equation}
(\mathbf{p}_{c}-\mathbf{p}_{h})\cdot E_{c}>0.
\end{equation}
Repeating the same procedure as before we get a necessary refrigerator
condition:
\begin{equation}
\mc P_{ch}\ge e^{-S_{c}}.
\end{equation}
In contrast to other measures like fidelity, $\mc P_{ch}$ has a very
simple statistical interpretation which also make it very easy to
measure in practice. $\sum_{i}p_{c,i}p_{h,i}$ is the probability
that two particles chosen from different baths will be in the same
state (like getting the same result with two different unbalanced
dices). To evaluate it, there is no need to keep track of the exact
result and then to estimate the probabilities $p_{c,i}$ and $p_{h,i}$
through their frequencies. One only needs to keep record if the results
are the same or not. Thus, $\mc P_{ch}$ corresponds to a binomial
random variable. Like the purity $\mc P_{ch}$ also satisfies $\mc P_{ch}\ge1/N$
(if the levels don't cross) but this is true for any two monotonic
distributions and does not give any indication of the heat machine's
functionality.

\section{Upper bounds on work and efficiency}

\subsection{Bounds on the maximal work production}

The first work upper bound can be obtained from (\ref{eq: W engine cond})
where the inequality $\ln\mathbf{p}_{c}\cdot\mathbf{p}_{h}\le\frac{1}{2}\ln\mc P_{c}\mc P_{h}$
is used to get to bring it to the local form: 

\begin{eqnarray}
\left\langle W\right\rangle  & \le & \frac{xR}{2-xR}[T_{h}S_{h}+T_{c}S_{c}+\frac{T_{h}+T_{c}}{2}\ln\mc P_{c}\mc P_{h}].\label{eq: W bound using F}
\end{eqnarray}
A different bound can be obtained by writing the work in a different
form:
\begin{equation}
\left\langle W\right\rangle =\frac{xR}{2-xR}[(T_{h}-T_{c})(S_{h}-S_{c})-T_{c}D_{KL}(\mathbf{p}_{h}|\mathbf{p}_{c})-T_{h}D_{KL}(\mathbf{p}_{c}|\mathbf{p}_{h})].\label{eq: W with D}
\end{equation}
The last two terms are always negative (including the minus sign).
Therefore the first term must be positive for the machine to perform
as an engine. $S_{h}\ge S_{c}$ is expected for engines as engine
transfer entropy from the hot bath to the cold bath but (\ref{eq: W with D})
quantifies how much large the entropy difference must be:
\begin{equation}
S_{h}-S_{c}\ge\frac{T_{c}D_{KL}(\mathbf{p}_{h}|\mathbf{p}_{c})+T_{h}D_{KL}(\mathbf{p}_{c}|\mathbf{p}_{h})}{T_{h}-T_{c}}.\label{eq: dS cond with D}
\end{equation}
The work expression (\ref{eq: W with D}) contain contain the non-local
quantity $D_{KL}$. To obtain a non-local upper bound we use the inequality:
\begin{eqnarray}
D_{KL}(\mathbf{p}_{c}|\mathbf{p}_{h}) & \ge & \frac{1}{2}(\sum\left|\mathbf{p}_{c,i}-\mathbf{p}_{h,i}\right|)^{2}\nonumber \\
 & \ge & \frac{1}{2}(\left|\mathbf{p}_{c}\right|^{2}+\left|\mathbf{p}_{h}\right|^{2}-2\left|\mathbf{p}_{c}\right|\left|\mathbf{p}_{h}\right|)\nonumber \\
 & = & \frac{1}{2}(\sqrt{\mc P_{c}}-\sqrt{\mc P_{h}})^{2},
\end{eqnarray}
and get:
\[
\]
\begin{equation}
\left\langle W\right\rangle \le\frac{xR}{2-xR}[(T_{h}-T_{c})(S_{h}-S_{c})-\frac{T_{c}+T_{h}}{2}(\sqrt{\mc P_{c}}-\sqrt{\mc P_{h}})^{2}].
\end{equation}
Different approaches and different inequalities may lead to work upper
bound that may perform better in certain regimes. In addition, further
reasonable restrictions on the system may lead to better results.
In appendix III we derive another work bound under some assumptions
about the energy levels structure.

\subsection{Upper bounds on the efficiency\label{sub: efficiency bounds}}

In the two level case the efficiency is simply given by $1-\frac{E_{2}^{c}-E_{1}^{c}}{E_{2}^{h}-E_{1}^{h}}$
and the population change cancels out. In the multilevel engine the
expression is more complicated:
\begin{equation}
\eta=1-\frac{\left\langle d\mathbf{p}\cdot\mathbf{E}_{c}\right\rangle }{\left\langle d\mathbf{p}\cdot\mathbf{E}_{h}\right\rangle }.
\end{equation}
Using the Clausius number defined in Sec. :
\begin{equation}
\eta=1-\frac{T_{c}}{T_{h}}-\frac{T_{c}}{\left|d\mathbf{p}\cdot\mathbf{E}_{h}\right|}\mc R.
\end{equation}
This result is still exact. To get a simpler bound we use again the
inequality $KL(\mathbf{p}_{c}|\mathbf{p}_{h})\ge\frac{1}{2}\left|\mathbf{p}_{c}-\mathbf{p}_{h}\right|^{2}$
and Cauchy-Schwarz in the numerator to obtain:
\begin{equation}
\eta<1-\frac{T_{c}}{T_{h}}-\frac{T_{c}}{\left|\mc E_{h}\right|}\left|\mathbf{p}_{c}-\mathbf{p}_{h}\right|,\label{eq: eta L2 L1 bound}
\end{equation}
where we used $\mc E_{h}$ to denote the centered energy vector:
\begin{equation}
\mc E{}_{h,i}=E_{h,i}-\frac{1}{N}\sum_{k=1}^{N}E_{h,k}.
\end{equation}
The replacement $\mathbf{E}_{h}\to\mc E_{h}$ is highly important.
It keeps the bound invariant to energy shifts and at the same time
makes the bound tighter. 

\begin{eqnarray}
\eta & \le & 1-\frac{T_{c}}{T_{h}}-\frac{T_{c}}{\left|\mc E_{h}\right|}\sqrt{\mc P_{c}+\mc P_{h}-2\mc P_{ch}}.\label{eq: eta L2 only}
\end{eqnarray}
In order to obtain a local-quantities bound we use the same procedure
we use $\mc P_{ch}\le\sqrt{\mc P_{c}}\sqrt{\mc P_{h}}$ and obtain
a weaker yet local upper bound :
\begin{equation}
\eta\le1-\frac{T_{c}}{T_{h}}-\frac{T_{c}}{\left|\mc E_{h}\right|}\left|\sqrt{\mc P_{c}}-\sqrt{\mc P_{h}}\right|.\label{eq: eta purity bound}
\end{equation}
This inequality is often very close to the Carnot efficiency. Yet,
it still shows that the baths purity imposes some limitations on the
efficiency. 

A different upper bound can be written down in term of the Wootters
statistical distance \cite{wootters81} between the hot and cold probability
distribution:
\begin{equation}
\mc L_{w}=\arccos(\sum_{i=1}^{N}\sqrt{p_{c,i}}\sqrt{p_{h,i}}).
\end{equation}
From \cite{DeffnerLutz2010generalizedClausius} it follows that:
\begin{equation}
J\ge\frac{16}{\pi^{2}}\mc L_{w}^{2},\label{eq: J Wooters}
\end{equation}
The efficiency bound obtained from using (\ref{eq: J Wooters}) is
not always smaller or larger than (\ref{eq: eta L2 L1 bound}) and
it uses the non-local quantity $\sum_{i=1}^{N}\sqrt{p_{c,i}}\sqrt{p_{h,i}}$
(Fidelity). Yet, it introduces another relation between the difference
in the statistics of the baths and the efficiency.

\[
\]
\[
\]

\section{The Ultra hot baths and the quasi-static regime\label{sub: Ultra hot regime}}

\subsection{The Ultra hot baths regime}

In atomic physics often the cold bath temperature is more difficult
to produce. In contrast it is fairly simple to produce a heat source
so that, roughly speaking, $E_{h}/T_{h},E_{c}/T_{c}\ll1$ (a more
rigorous condition will be given later) . In this section we study
the multilevel heat engine operation when the hot bath is so hot that
we consider only the first order correction in $1/T_{b}$ to the $T_{b}\to\infty$
limit where the $p_{b,i}\to1/N$. The ultra-hot regime is defined
by the small parameters: 
\begin{equation}
\frac{\Delta E_{c}}{T_{c}},\frac{\Delta E_{h}}{T_{h}}\ll1,
\end{equation}
where the $\Delta E_{b}=E_{b,max}-E_{b,min}$ is the gap of the bath
$ $

First order expansion in $\{\beta_{c}=1/T_{c},\beta_{h}=1/T_{h}\}$
yields:

\begin{eqnarray}
W^{ultra\: hot} & = & \frac{xR}{2-xR}\frac{1}{N}[(\beta_{c}+\beta_{h})\mc E_{c}\cdot\mc E_{h}-\beta_{c}\left|\mc E_{c}\right|^{2}-\beta_{h}\left|\mc E_{h}\right|^{2}].\label{eq: W ultra hot}
\end{eqnarray}
This time the centered energy form $\mc E_{b}$ emerges naturally
from the free energy normalization factor in the ultra-hot limit.
$W>0$ yields the necessary and sufficient engine condition:

\begin{eqnarray}
\mc E_{c}\cdot\mc E_{h} & > & \frac{\beta_{c}\left|\mc E_{c}\right|^{2}+\beta_{h}\left|\mc E_{h}\right|^{2}}{\beta_{c}+\beta_{h}},
\end{eqnarray}
Applying the Cauchy-Schwartz we get a \textit{necessary} condition
for ultra-hot swap engines:
\begin{equation}
1<\frac{\left|\mc E_{h}\right|}{\left|\mc E_{c}\right|}<\frac{T_{h}}{T_{c}},\label{eq: eh/ec ultra}
\end{equation}

To optimize the work output, let us first assume that the norms of
the energies $\left|\mc E_{c}\right|$ and $\left|\mc E_{h}\right|$
are fixed. Under these two constraints the last two terms in (\ref{eq: W ultra hot})
are fixed. To maximize the first term, and consequently the work,
$\mc E_{c}$ must be to be parallel to $\mc E_{h}$:
\begin{eqnarray}
\mc E_{h,i} & = & \mc C\mc E_{c,i},\label{eq: zeta def}\\
1 & \le\mc C\le & \frac{T_{h}}{T_{c}},\label{eq: zeta range}
\end{eqnarray}
where $\mc C$ is the compression ratio and (\ref{eq: zeta range})
follows from (\ref{eq: eh/ec ultra})$ $. For ``uniform compression''
(\ref{eq: zeta def}) the device works as a refrigerator when $\mc C<1$,
and as a heater when $\mc C\ge T_{h}/T_{c}$. In the regime (\ref{eq: zeta range})
it performs as an engine. The uniform compression can be studied for
any temperature but in the ultra hot regime it is found that the uniform
compression maximizes the work when the energy norms are fixed. Now
we remove the restriction that $\left|\mc E_{h}\right|$ is fixed
and instead use (\ref{eq: zeta def}) in (\ref{eq: W ultra hot}).
Imposing $ $$\partial_{\mc C}W=0$ we get: $ $ 

\begin{equation}
\mc C_{max,\left|\mc E_{h}\right|}=\frac{T_{h}}{\frac{1}{2}(T_{h}+T_{c})}\le2,
\end{equation}
where $\left|\mc E_{h}\right|$ in the subscript signifies the $\left|\mc E_{h}\right|=const$
optimization constraint. The efficiency is:

\begin{equation}
\eta_{max,\left|\mc E_{h}\right|}^{ultra\: hot}=1-1/\mc C_{max,\left|\mc E_{h}\right|}=\frac{T_{h}-T_{c}}{2T_{h}}=\frac{1}{2}\eta_{c}\le\frac{1}{2}.
\end{equation}
Where $\eta_{c}$ is the Carnot efficiency. We note that this efficiency
is always tighter (smaller) than the Novikov, Curzon and Ahlborn (NCA)
efficiency bound $\eta_{CA}=1-\sqrt{T_{c}/T_{h}}$\cite{novikov1958efficiency,curzon75}.
The NCA derivation assumption assumes the working substance is in
some well-defined temperature. As Eqs. (\ref{eq: rho C steady}) and
(\ref{eq: rho A steady}) indicate this is not true in our model (unless
$xR=1$). Furthermore, in our model the heat flow rate is not proportional
to the temperature difference as assumed in the NCA model (the Newton
heat transfer law). So, although our result is consistent with NCA,
we claim there is no clear reason why the two should be compared.
If we require that $\left|\mc E_{c}\right|$ is the constraint instead
of $\left|\mc E_{h}\right|$, then we get $\mc C_{max,\left|\mc E_{h}\right|}=\frac{\frac{1}{2}(T_{h}+T_{c})}{T_{c}}$
and $\eta_{max,\left|\mc E_{c}\right|}^{ultra\: hot}=\frac{T_{h}-T_{c}}{T_{h}+T_{c}}=\frac{\eta_{c}}{2-\eta_{c}}$.
Interestingly, this efficiency is always larger than the NCA result.
The same expression for the efficiency was obtained in \cite{GelbwaserAlickiKurizkiMinUniEngine2010}
for a different type of engine in the ultra hot regime.

The expression for maximal work for a given $\left|\mc E_{h}\right|$
is: 
\begin{equation}
W_{max}^{ultra\: hot}=\frac{xR}{2-xR}\frac{1}{N}\frac{(T_{h}-T_{c})^{2}}{4T_{c}T_{h}^{2}}\mc E_{h}^{2},
\end{equation}
The expression looks asymmetric in $T_{c},T_{h}$ and $\mc E_{c},\mc E_{h}$
since we already applied the optimal ratio between $\mc E_{c}$ and
$\mc E_{h}$. Bringing back $\mc E_{c}$ to the equation will restore
symmetry but will hide the choice already made for $\mc E_{c}$ by
the optimization procedure. At first sight, it seems that the work
becomes smaller when the number of levels increases. Yet, $\mc E_{h}^{2}$
may also depend on $N$. For example if the levels are degenerate
so that there are $N/2$ replicas of two levels, then we get that
the work does not depend on the number of replicas, as expected.

It is interesting to look on the Clausius number in this limit:
\begin{eqnarray}
\mc R^{ultra\: hot} & = & \frac{xR}{2-xR}\sum_{i}(\frac{\mc E_{h,i}}{T_{h}}-\frac{\mc E_{c,i}}{T_{c}})(\frac{E_{h,i}}{T_{h}}-\frac{E_{c,i}}{T_{c}})=\nonumber \\
 & = & \frac{xR}{2-xR}\sum_{i}(\frac{\mc E_{h,i}}{T_{h}}-\frac{\mc E_{c,i}}{T_{c}})(\frac{\mc E_{h,i}}{T_{h}}-\frac{\mc E_{c,i}}{T_{c}}+const)
\end{eqnarray}
The constant term can be dropped out since it contributes zero to
the total sum. After the constant term is removed, all the terms in
the Clausius sum are positive. This is not true for below the ultra
hot baths regime. In this regime, the Clausius inequality holds since
the probability difference and the Clausius factor $\frac{\mc E_{h,i}}{T_{h}}-\frac{\mc E_{c,i}}{T_{c}}$
have the exact same form.

\subsection{Almost ``Quasi-static evolution'' at finite time}

Consider the case where the swap parameter is small enough so that
after a collision the change in a bath particle population, $dp_{b,i}$,
is small with respect to the original bath population so that $\delta p_{b,i}\ll p_{b,i}$.
At the end of appendix I it is shown that the energy diagonal form
of the density matrices is conserved in an energy population swap
(this trivially holds in a density swap interaction). Consequently,
it is possible to use just $\mathbf{p}_{b}$ and $\delta\mathbf{p}_{b}$
to calculate the von Neumann entropy change. To first order in $\delta p_{b,i}$
the change in the \textit{bath} particle's entropy is given by: 
\begin{equation}
\left\langle \delta S_{b}\right\rangle =S(\vec{p}_{b}+\delta\vec{p}_{b})-S(\vec{p}_{b})=\left\langle \delta p_{b}\right\rangle \frac{\Delta E_{b}}{T_{b}}=\frac{\left\langle \delta Q_{b}\right\rangle }{T_{b}}\label{eq: delta S}
\end{equation}

Note that this equation holds only for the bath particles and for
tiny deviations from the Gibbs state. The engine is not in a thermal
state at all, and cannot be assigned a temperature. However, in a
complete cycle, on average, the engine returns to its initial state
so the engine does not contribute to the average total entropy production
of the total system. Therefore the total entropy increase in both
baths and the system satisfies:
\begin{equation}
\left\langle \delta S_{h}\right\rangle +\left\langle \delta S_{c}\right\rangle \ge0,
\end{equation}
by virtue of the Clausius inequality for swap heat machines proven
earlier. 

Assuming that on average there are $n$ cold collisions and $n$ hot
collision, the work can be expressed in terms of the entropy changes:
\begin{equation}
\left\langle W\right\rangle =nT_{h}\left\langle \delta S_{h}\right\rangle +nT_{c}\left\langle \delta S_{c}\right\rangle 
\end{equation}

Although (\ref{eq: delta S}) seems like a plausible property for
a bath, it is not mandatory and in our model it emerges only in the
weak coupling limit (small population change per collision). 

Even though we assumed the change in a single collision is small it
does not mean that the heat exchange in strokes $B$ or $C$ must
be small in our model. If we allow multiple collisions, it is possible
to have large changes while still satisfying $\Delta Q_{b}=T_{b}\Delta S_{b}=T_{b}n\delta S_{b}$. 

In our entropy considerations we did not take into account mixing
entropy so it is not claimed that (\ref{eq: delta S}) is the change
in the entropy of the system but just the part of entropy change that
is responsible for the heat exchange. 

\[
\]

\section{Conclusion}

We have presented an analysis of a multilevel heat machine that is
driven by a partial swap interaction sequentially with a hot and cold
baths. Our approach emphasizes the back reaction of the engine on
the baths. By studying the bath's thermodynamics observables we were
able to relate the work, efficiency and entropy production to a number
of measurable bath properties like purity and mutual coincidence.
The equivalent of Clausius inequality in this system was generalized
to higher order energy moments. We identified a quasi-static regime
in finite time evolution and an ultra-hot regime where stronger statements
can be made. 

While our findings are valid for any multilevel engine and collision
that can bring the system to a Gibbs state, it is interesting to consider
other bath models like the ones used in continuous engines where different
levels interact with different baths or where the thermalization rate
is different for different levels.

\[
\]

\section*{Appendix 1 - The two-level swap Hamiltonian }

The partial swap unitary is given by: 
\begin{equation}
U=e^{-i\frac{1}{2}\phi\sum_{i}\sigma_{i}\otimes\sigma_{i}}\label{eq: s x s swap}
\end{equation}
where $\sigma_{i}$ are the Pauli matrices and $\phi$ is the swap
angle for $\phi=\pi/2$ a complete swap takes place. Let $\rho_{1,2}$
be diagonal matrices. The reduced density matrix of particle 1after
applying U is: 
\begin{equation}
\rho_{1}'=tr_{2}(U\rho_{1}\otimes\rho_{2}U^{\dagger})
\end{equation}

One can verify that the density swap rule is satisfied:
\[
\rho_{1}'=\cos^{2}\phi\rho_{1}+\sin^{2}\phi\rho_{2}
\]

if $\rho_{1,2}$ are diagonal then $\rho_{1,2}'$ are also diagonal. 

Since the Hamiltonian is written in terms of the Pauli matrices it
is not straightforward to generalize it to a multilevel swap operation. 

For energy population swap, the generalization is simpler when the
two-particle states are considered. A complete swap operation yields:
$U_{cs}\ket{i,j}=U_{cs}\ket{j,i}$. The states $\ket{i,i}$ are invariant
under this operation. The complete swap unitary in the two-particle
state can be written as:
\begin{equation}
\begin{array}{c}
11\\
22\\
33\\
12\\
21\\
13\\
31\\
\vdots
\end{array}\left(\begin{array}{ccccccc}
1\\
 & 1\\
 &  & ..\\
 &  &  &  & 1\\
 &  &  & 1\\
 &  &  &  &  &  & 1\\
 &  &  &  &  & 1\\
\vdots
\end{array}....\right)\label{eq: block U}
\end{equation}

The column to the left shows how the states are ordered in the two-particle
density matrix. In this form the unitary has a clear block diagonal
structure. In space of each non identical state $\{\ket{ij},\ket{ji}\}$
a simple spin flip takes place:
\begin{equation}
a\ket{ij}+b\ket{ji}\to a\ket{ji}+b\ket{ij},
\end{equation}
Replacing the spin flip by a more general rotation in this Hilbert
subspace, it is easy to deduce the partial swap time-independent Hamiltonian:

\begin{equation}
H=\sum_{ij}\phi_{ij}\ketbra{ij}{ji}.
\end{equation}
The $\phi_{ii}$ terms just contribute a phase to the invariant states.
A complete swap takes place when $\phi_{i,j\neq i}=\pm\pi/2$. 

In general, there is no particular reason why the rotation rate, $\phi_{ij}$
should be the same for all pairs of states. However, when it is the
same, $\phi_{i\ne j}=\phi$ can show that the energy population swap
probability swap rule follows:
\begin{equation}
p_{1}'=diag(tr_{2}(U\rho_{1}\otimes\rho_{2}U^{\dagger}))=\cos^{2}\phi p_{1}+\sin^{2}\phi p_{2}
\end{equation}
This swap model is specially designed for the energy basis. Consequently,
for partial swap, in general, it does not satisfy the ``density swap''
rule but the energy basis probability rule. Yet, a proper choice of
$\phi_{ii}$ leads to the density swap Hamiltonian that appears in
the exponent of (\ref{eq: s x s swap}). We conclude by noting the
important fact that the energy population Hamiltonian preserves the
energy diagonal form after the partial trace. That is, if the input
state is $\rho_{1}\otimes\rho_{2}$ where $\rho_{1},\rho_{2}$ are
diagonal in the energy basis then $ $$\rho_{1(2)}'=tr_{2(1)}(U\rho_{1}\otimes\rho_{2}U^{\dagger})$
will be diagonal in the energy basis. For simplicity we illustrate
it for the $\phi_{i\ne j}=\phi$ case. The density matrix after the
collision and before the trace is:
\begin{equation}
U\rho_{1}\otimes\rho_{2}U^{\dagger}=\cos^{2}\phi\rho_{1}\otimes\rho_{2}+\sin^{2}\phi\rho_{2}\otimes\rho_{1}+\sum_{i\neq j}f_{ij}\ketbra{ij}{ji}
\end{equation}
where $f_{ij}$ are some complex coefficients. The third term vanishes
when taking partial trace on either particle. Due to the block structure
(\ref{eq: block U}) this holds even when $\phi_{i\ne j}$ depends
on $i$ and $j$. 
\[
\]

\section*{Appendix II - Markovian swap formulation}

As shown in Sec. \ref{sec: evo fluct} in steady state all coherences
of the density matrix vanish thus only the diagonal elements are important
and the Markov chain formalism can be used to describe the evolution
of the diagonal elements (probabilities). In this appendix it will
be more convenient to use Dirac's ``Bra-Ket'' notation for the probabilities
vectors in order to distinguish between right vectors and left vectors.
However the normalization is still the regular probability normalization.

The probabilities before and after the collision with the hot bath
satisfy:
\begin{eqnarray}
\ket{p_{s}'} & = & K_{h}\ket{p_{s}},\\
K_{h} & =\tilde{x} & \ketbra{p_{T_{h}}}{1,1,1..}+(1-\tilde{x})I_{N\times N},
\end{eqnarray}

where $\tilde{x}$ is an effective swap parameter that may contain
``classical'' contribution from the collision probability R and
quantum contribution from a quantum partial swap $(x<1\: or\:\phi<\pi/2)$.
$I_{N\times N}$ is the identity operator. One can verify that the
above equations lead to the population swap rule $p_{s}'=(1-\tilde{x})p_{s}+\tilde{x}p_{h}$.
A Markov chain with a single steady state vector is always associated
with a left eigenvector of the form $\bra{1,1,1..}$. Thus, $\ket{p_{T_{h}}}$
is a right eigenvector of $K_{h}$ and it has a unity eigenvalue:
$K_{h}\ket{p_{T_{h}}}=\ket{p_{T_{h}}}$. 

The engine operation over one cycle starting form stroke $A$ is given
by:
\begin{eqnarray}
K_{A\to A} & = & K_{c}K_{h}\nonumber \\
 & = & [\tilde{x}^{2})\ket{p_{T_{c}}}+\tilde{x}(1-\tilde{x})\ket{p_{T_{c}}}+\tilde{x}(1-\tilde{x})\ket{p_{T_{h}}}]\bra{1,1,1..}\nonumber \\
 & + & (1-\tilde{x})^{2}I_{N\times N}.
\end{eqnarray}
This operator has a clear interpretation. The $\tilde{x}^{2}\ket{p_{T_{c}}}$
describe two complete swaps that occur with probability $\tilde{x}^{2}$.
Since both swap events are complete, the population is determined
by the last swap with the cold bath. The $\tilde{x}(1-\tilde{x})$
terms describe the probability for a single complete swap event and
the $(1-\tilde{x})^{2}$ term describes a zero swap event. 

Clearly the invariant steady state is $(x(1-x)+x^{2})\ket{p_{T_{c}}}+x(1-x)\ket{p_{T_{h}}}$.
$ $ After normalization we obtain the steady state eigenvector:
\begin{equation}
\ket{p_{e}^{A}}=\frac{\ket{p_{T_{c}}}+\ket{p_{T_{h}}}-\tilde{x}\ket{p_{T_{h}}}}{2-\tilde{x}}\label{eq: pc mark}
\end{equation}

Repeating this for a cycle starting from stroke $C$ and using $K_{C\to C}=K_{c}K_{h}$
one gets:
\begin{equation}
\ket{p_{e}^{C}}=\frac{\ket{p_{T_{c}}}+\ket{p_{T_{h}}}-\tilde{x}\ket{p_{T_{c}}}}{2-\tilde{x}}\label{eq: ph mark}
\end{equation}

Both (\ref{eq: pc mark}) and (\ref{eq: ph mark}) agree with the
steady state population obtained in Sec. (\ref{sec: evo fluct}),
when replacing $\tilde{x}\to xR$

\section*{Appendix III - Alternative work upper bound}

The expression for the work is a standard inner product over the real
vectors. By using the Cauchy-Schwarz inequality we get:

\begin{equation}
\left|\left\langle W\right\rangle \right|\le\frac{xR}{2-xR}\left|\mathbf{p}_{c}-\mathbf{p}_{h}\right|\left|\mathbf{\mc E}_{c}-\mathbf{\mc E}_{h}\right|\label{eq: gen W upper bound}
\end{equation}
At first, it seems, that this separation between statistics and energy
is not very useful. If all $\vec{p}_{c}$ and $\vec{p}_{h}$ have
to be known, then there is not much difference from calculating the
exact value of the work. However, $\left|\mathbf{p}_{c}-\mathbf{p}_{h}\right|$,
can be expressed in term of simple scalars quantities that characterize
the baths.
\begin{equation}
\left|\mathbf{p}_{c}-\mathbf{p}_{h}\right|^{2}=\left|\mathbf{p}_{c}\right|^{2}+\left|\mathbf{p}_{h}\right|^{2}-2\mathbf{p}_{c}\cdot\mathbf{p}_{h}=\mc P_{c}+\mc P_{h}-2\mc P_{ch}
\end{equation}

The upper bound (\ref{eq: gen W upper bound}) holds for any diagonal
distribution and even if the distribution is not known exactly it
is enough to measure $\mc P_{c},\mc P_{h}$ and $\mc P_{ch}$ to evaluate
$\left|\mathbf{p}_{c}-\mathbf{p}_{h}\right|$. Since $\mathbf{p}_{c}$
and $\mbox{p}_{h}$ are ordered if there are no level crossings, we
can use Chebyshev's sum inequality and get: $\mathbf{p}_{c}\cdot\mathbf{p}_{h}\ge1/N$
so that:

\begin{eqnarray}
\left|\left\langle W\right\rangle \right| & \le & \frac{xR}{2-xR}\sqrt{\mc P_{c}+\mc P_{h}-2\mc P_{ch}}\left|\mathbf{\mc E}_{c}-\mathbf{\mc E}_{h}\right|\label{eq: W bound CS}\\
 & \le\frac{xR}{2-xR} & \sqrt{\mc P_{c}+\mc P_{h}-\frac{2}{N}}\left|\mathbf{\mc E}_{c}-\mathbf{\mc E}_{h}\right|\label{eq: W bound CS 1/N}
\end{eqnarray}
Note that: $\sqrt{\mc P_{c}+\mc P_{h}-\frac{2}{N}}\le\sqrt{2}\sqrt{\frac{N-1}{N}}$.
If all the levels are compressed (not necessarily by the same factor)
so that: 
\begin{eqnarray}
E_{c,i} & = & \frac{1}{\mc C_{i}}E_{h,i}\\
\mc C_{i} & \ge & 1\label{eq: zeta cond}
\end{eqnarray}
In this case, we can use $(a-b)^{2}\le\left|a-b\right|\left|a+b\right|=\left|a^{2}-b^{2}\right|\:\forall\: ab>0$
to get:

Finally we can write down a bound based only on individual bath properties:

\begin{eqnarray}
\left|\left\langle W\right\rangle _{\mc{\mc C}_{i}\ge1}\right| & \le & \frac{xR}{2-xR}\sqrt{\mc P_{c}+\mc P_{h}-\frac{2}{N}}\left|\mathbf{E}_{h}-\mathbf{E}_{c}\right|\\
 & \le & \frac{xR}{2-xR}\sqrt{\mc P_{c}+\mc P_{h}-\frac{2}{N}}\sqrt{\mathbf{E}_{h}^{2}-\mathbf{E}_{c}^{2}}\label{eq: W< diff(|Eh|^2)}
\end{eqnarray}
We cannot use $\epsilon$ here, as it will typically violate (\ref{eq: zeta cond}).
Furthermore, although (\ref{eq: W< diff(|Eh|^2)}) seems to imply
that $W\to0$ if $\left|\mathbf{E}_{h}\right|=\left|\mathbf{E}_{c}\right|$,
the norms cannot be equal and satisfy $\mc C_{i}\ge1$ (with the exception
of the uninteresting case where all the compression factors $\mc C_{i}$
are equal to one in which the work is trivially equal to zero). The
advantage of bound (\ref{eq: W< diff(|Eh|^2)}) is that it uses only
local scalars.

\ack{}{}

Work supported by the Israel Science Foundation. Part of this work
was supported by the COST Action MP1209 'Thermodynamics in the quantum
regime'

\section*{References}

\bibliographystyle{iopart-num}
\bibliography{dephc2}
 
\end{document}